# IoT and Predictive Maintenance in Industrial Engineering: A Data-Driven Approach

**Dr. P.Vijaya Bharati[1], Dr. J.S.V.Siva Kumar[2], Mr. Sathish Kumar Anumula[3], Mr. P Vamshi Krishna[4], Dr. Sangam Malla[5]**

[1]Associate Professor, Department of CSE, Vignans Institute of Engineering for Women,Visakhapatnam, Andhrapradesh-530046.

Email ID: pvijayabharati@gmail.com

[2]Associate Professor, Department of EEE, GMRIT, Rajam, Andhra Pradesh - 532127

Email ID: sivakumar.jsv@gmrit.edu.in

[3]Thorrur, Thurkamjal, Hyderabad, Telangana - 501511

Email ID: sathishkrishna@gmail.com

[4]Assistant Professor, Department of Freshmen Engineering, St. Martin's Engineering College, Dhulapally, Secunderabad-500100

Email ID: vamshi.pearala@gmail.co

[5]Assistant Professor, Department of Computer Science, Udayanath Autonomous College of Science and Technology,Adaspur,Cuttack,Odisha-754011

Email ID: sangam.malla2015@gmail.com

**ABSTRACT**

Fourth Industrial Revolution has brought in a new era of smart manufacturing, wherein, application of Internet of Things (IoT), and data-driven methodologies is revolutionizing the conventional maintenance. With the help of real-time data from the IoT and machine learning algorithms, predictive maintenance (PdM) allows industrial systems to predict failures and optimize machines' life. This paper presents the synergy between the Internet of Things and predictive maintenance in industrial engineering with an emphasis on the technologies, methodologies, as well as data analytics techniques, that constitute the integration. A systematic collection, processing, and predictive modeling of data is discussed. The outcomes emphasize greater operational efficiency, decreased downtime, and cost-saving, which makes a good argument as to why predictive maintenance should be implemented in contemporary industries.

***Keywords:*** *IoT, Predictive Maintenance, Industrial Engineering, Machine Learning, Data Analytics, Smart Manufacturing, Industry 4.0, Condition Monitoring, Preventive Maintenance, Fault Detection*

## 1. INTRODUCTION

Reduction in the rate at which digital technologies are changing has caused major alterations to occur within the industrial sector. One of the most revolutionary evolutions is the adoption of the Internet of Things (IoT) into the industrial engineering disciplines [1]. IoT, through its network of interlinked sensors, actuators and systems, enables real time collection, monitoring and control of data in manufacturing operations. Applications of IoT in industrial settings are becoming more promising by the day and one of the most promising of these is with regard to Predictive Maintenance (PdM) in the industrial sphere —a methodology that allows companies to make predictions based on data-driven insights into equipment failures before they have a chance to occur. Such forward thinking guarantees best results in machines performance and minimizes unforeseen downtimes for higher return of operational efficiency.

Industries have in the past used corrective or preventive maintenance strategies. Corrective maintenance dictates repairing equipment once it has failed, leading to expensive breakdown times, and lost production. Preventive maintenance, however, fixes maintenance activities into time intervals regardless of condition of the equipment. Though preventive maintenance is more efficient than reactive maintenance processes, it can still result in unnecessarily servicing or failure omission. Such constraints have given rise to the development of the predictive maintenance which uses sensor data, statistical analysis and machine learning algorithms to determine the state of the equipment and how long it should be maintained [14-15].

This concept of predictive maintenance is not all new. however, due to the transformation of the IoT and big data technologies, its effectiveness has been greatly elevated. IoT sensors are currently able to collect massive operational data – temperature, pressure, vibrations, acoustics and level of energy consumption – in real time. When processed and analyzed as they should be, these data streams provide a lot of insight into industrial assets' health. Moreover, the decision to implement cloud computing and edge analytics will make sure this data can be stored, processed and visualized in a scalable and timely manner.

Amid this changing scenery, data-based methods for predictive maintenance are finding shelf space. Such methods commonly use the ML and DL models to identify anomalies, RUL estimation of components, and recommendation development of service. The models are trained on historical equipment data, and adjusted iteratively based on continuous learning from real-time inputs. Such methods for analyzing time-series data and identifying patterns suggestive of upcoming failure as Random Forest, Support Vector Machines (SVM), and Recurrent Neural Networks (RNN's), especially the Long Short-Term Memory (LSTM) models are widely used [9-11].

The positives of the predictive maintenance are well documented in literature. When PdM strategies are adopted, the other organizations experience improved serviceability of equipment, better safety standards and huge cost saving. For instance, in the manufacturing, aerospace, and energy industries, predictive maintenance will result in cost savings in terms of maintenance by 25 to 30 percent and downtime by 45 percent. Notwithstanding such benefits, a large number of industries are yet at the early stages of IoT supported PdM solutions implementation with the majority of the cited challenges documented including those associated with data quality, system integration, cybersecurity, and the need for specialized analytical skillsets.

As a discipline, Industrial Engineering is ideally placed to be the leader of this transition. Through integration of systems optimization principles, operations management science, and data science, industrial engineers are capable of building and deploying complete PdM frameworks suitable for individuals' organizational requirements. This entails the choice of suitable IoT sensors, identification of key performance indicators (KPIs), designing strong predictive models and better integration of hardware and software components in the industrial ecosystem.

This paper describes the confluence between the IoT and predictive maintenance, as related fields in industrial engineering. It suggests an all-embracing data-driven construct consisting of sensor data acquisition, machine learning incorporating predictive modeling and cloud-based visualization tools. By means of a case study on real industrial equipment, the paper shows its applied advantages and technological feasibility. Furthermore, it discusses challenges in methodology and future implementation and research recommendations. This study seeks to add to the academic and practical knowledge of how these technologies can be synthetically utilized to produce smarter, more productive industrial ventures [12].

*Novelty and Contribution*

Novelty in this research work is positioned in its integrated, real-time predictive maintenance architecture that exploits IoT technologies and sophisticated machine learning algorithms within an industrial engineering regime. Whereas previous research has addressed separate elements of predictive maintenance, or have used theoretical models with historical data, this paper describes a practically oriented, end-to-end system that can be adopted straight away in operational industrial settings [3-5].

One major contribution of this work, therefore, is the development of a hybrid IoT-ML architecture that includes the use of edge computing for the initial handling of data and cloud platforms for extensive analytics and visualization. The double-layering design is scalable, low latent, and decision-making efficient for high throughput industrial situations.

The other unique feature is the use of longitudinal time-series data that are collected from real manufacturing equipment. Contrary to numerous academic works relying on synthetic or publicly available data, the paper at hand uses the genuine operational data for training, validation and performance analysis of different predictive models. This enables better performance comparison and provides clues on realistic deployment problems, including sensor noise, signal loss, and data scarcity.

Additionally, the paper presents the idea of multi-model prediction approach via Random Forest, SVM, and LSTM networks. This collection of models is examined comparatively in order to determine best application areas for every algorithm, for increased predictive accuracy and robustness. Such systematic appraisal is usually absent in previous literature which is primarily characterized by single model approaches.

Besides technical contributions, the paper also provides a systematic approach that industrial engineers should use to deliver IoT-based predictive keeps in their companies. This involves sensor selection, feature engineering, model selection criteria and integration of real-time monitoring. This work closes the gap between theory and industrial application making complex data science techniques accessible in a working process.

Finally, the presence of a live industrial case study provides extra credibility to the practical applicability of the work. Attributed reductions in downtime and maintenance costs are the practical demonstration of real effects of data-driven predictive maintenance systems, and, therefore, encourage wider usage in similar industrial settings.

Complementing each other, these elements form a robust contribution to the space of industrial engineering, smart manufacturing, and IoT-based predictive analytics, promoting knowledge and implementation strategies in practice and theory.

## 2. RELATED WORKS

In 2021 Dolgui et.al. and D. Ivanov et.al., [8] suggested the significant development of study in predictive maintenance as smart technologies and data-centric approaches emerge. The previous maintenance models were mostly based on the reactive and time-based approach that too often resulted in inefficiencies, higher costs, and unexpected failures of the machinery. With the arrival of condition monitoring systems and digital sensing technologies, industries started to move towards the more proactive approach. Such developments paved the way to the integration of PM into industrial operations.

Use of IoT has in recent years revolutionized the predictive maintenance landscape. IoT-enabled devices support regular monitoring of the important operational parameters, vibration, temperature, noise, and load. The numbers of real-time data generated by these sensors are huge, and they open up other possibilities with regard to advanced analysis and machine learning applications. Existing research revealed that integration of sensor based data collection with predictive algorithms improves detection of faults and scheduling of maintenance works.

Some of the industries that have been discussing the introduction of predictive maintenance include manufacturing, transportation, energy, and aerospace industries, with a view of enhancing assets utilization. In most cases, support vector machines, decision trees, and neural networks have been used as machine learning models to determine the early signs of equipment degradation. Recurrent neural networks have also put on momentum because they can model temporal order and discover complex failure patterns using time-series data.

Further studies have explored and uncovered the significance of the edge computing and cloud infrastructure in shaping predictive maintenance systems. Edge computing enables local processing of data close to their sources, minimizing latency and the bandwidth needs. Such a double computing structure allows for strong, responsive and expansive maintenance frameworks.

In 2021 Z. Xu et.al. and J. H. Saleh et.al., [13] proposed the notwithstanding the great strides, there are issues with the adoption of predictive maintenance at scale. Data heterogeneity, the absence of interoperability of IoT devices and the absence of generalization model across different types of machines are constant barriers. Besides, issues of data privacy and cybersecurity are still limiting wider adoption, especially in the critical infrastructure sectors.

According to the literature, user-friendly interfaces and decision-support systems are also of paramount importance to make sure that maintenance recommendations extracted from the predictive model can be operational and understood by the operational staff. Visualization dashboards, alert systems, and synchronization with enterprise asset management tools are quickly gaining the status of vital parts of the predictive maintenance ecosystems.

In 2021 Bousdekis et.al., K. Lepenioti et.al., D. Apostolou et.al., and G. Mentzas et.al., [2] introduced the studies demonstrate the necessity to combine the IoT, the machine learning, and the industrial engineering ideas to develop efficient, real-time, and scalable predictive maintenance systems. This paper extends these findings by offering a unifying and practical approach to solving both the technical and operational problems of deploying predictive maintenance in industrial contexts.

## 3. PROPOSED METHODOLOGY

This section presents a practical, modular framework for predictive maintenance in industrial environments using loT devices and data-driven analytics. The methodology consists of four main stages: Sensor Data Acquisition, Feature Engineering, Predictive Modeling, and Decision Support. Each stage integrates core mathematical foundations to ensure reliability and accuracy in maintenance predictions [6].

*A. Sensor Data Acquisition*

IoT sensors are installed on critical machine components to continuously monitor operational parameters such as vibration, temperature, acoustic signals, and current. These values are recorded over time and represented as multivariate time-series data $X(t)$, where:

$$X(t) = \begin{bmatrix} x_1(t) \\ x_2(t) \\ \vdots \\ x_n(t) \end{bmatrix}$$

To smooth the raw signals and reduce noise, a moving average filter is applied:

$$\bar{x}(t) = \frac{1}{k} \sum_{i=t-k+1}^{t} x(i)$$

where $k$ is the window size. For temperature, pressure, and vibration, these smoothed values are forwarded to the feature engineering module.

*B. Feature Engineering*

Feature vectors are extracted using statistical transformations from the time series. Standard deviation $\sigma$ and root mean square (RMS) are computed to capture variability and energy content:

$$\sigma = \sqrt{\frac{1}{N} \sum_{i=1}^{N} (x_i - \mu)^2}$$

$$\text{RMS} = \sqrt{\frac{1}{N} \sum_{i=1}^{N} x_i^2}$$

Spectral entropy is calculated to measure the signal's irregularity:

$$H = -\sum_{i=1}^{n} p_i \log_2 p_i$$

where $p_i$ is the power spectrum probability of component $i$. Another key indicator is the kurtosis $K$, used to detect faults such as bearing wear:

$$K = \frac{1}{N} \sum_{i=1}^{N} \left(\frac{x_i - \mu}{\sigma}\right)^4$$

These extracted features are fed into a supervised learning model for training and prediction.

*C. Predictive Modeling*

A hybrid model comprising Support Vector Machine (SVM) and Long Short-Term Memory (LSTM) networks is proposed. For binary classification of failure conditions using SVM, the decision boundary is defined as:

$$f(x) = \text{sign}(w^T x + b)$$

Here, $w$ is the weight vector and $b$ the bias term. The optimization problem minimizes the following cost function:

$$\min_{w,b} \frac{1}{2} \|w\|^2 + C \sum_{i=1}^{n} \xi_i$$

subject to:

$$y_i(w^T x_i + b) \geq 1 - \xi_i, \xi_i \geq 0$$

To model time dependencies and predict Remaining Useful Life (RUL), the LSTM network uses a gated architecture. The memory cell update equation is:

$$c_t = f_t \cdot c_{t-1} + i_t \cdot \tilde{c}_t$$

where:

- $f_t$ is the forget gate,
- $i_t$ is the input gate,
- $\tilde{c}_t$ is the candidate state.

The final predicted RUL $\hat{R}(t)$ is modeled as:

$$\hat{R}(t) = \beta_0 + \sum_{i=1}^{n} \beta_i \cdot F_i(t)$$

where $F_i(t)$ are the extracted features and $\beta_i$ the regression coefficients learned from training.

*D. Decision Support System*

The maintenance decision is made when the predicted RUL drops below a threshold $\tau$. Let the alert function be:

$$\text{Alert }(t) = \begin{cases} 1 & \text{if } \hat{R}(t) \leq \tau \\ 0 & \text{otherwise} \end{cases}$$

This binary function ensures that maintenance is only triggered when truly necessary. Additionally, for costbenefit analysis, a savings function is introduced:

$$S = C_p - \left(C_f + C_d\right)$$

where:

- $C_p$ is the cost of predictive maintenance,
- $C_f$ is failure-related repair cost,
- $C_d$ is downtime cost.

If $S > 0$, the predictive model is deemed economically viable.

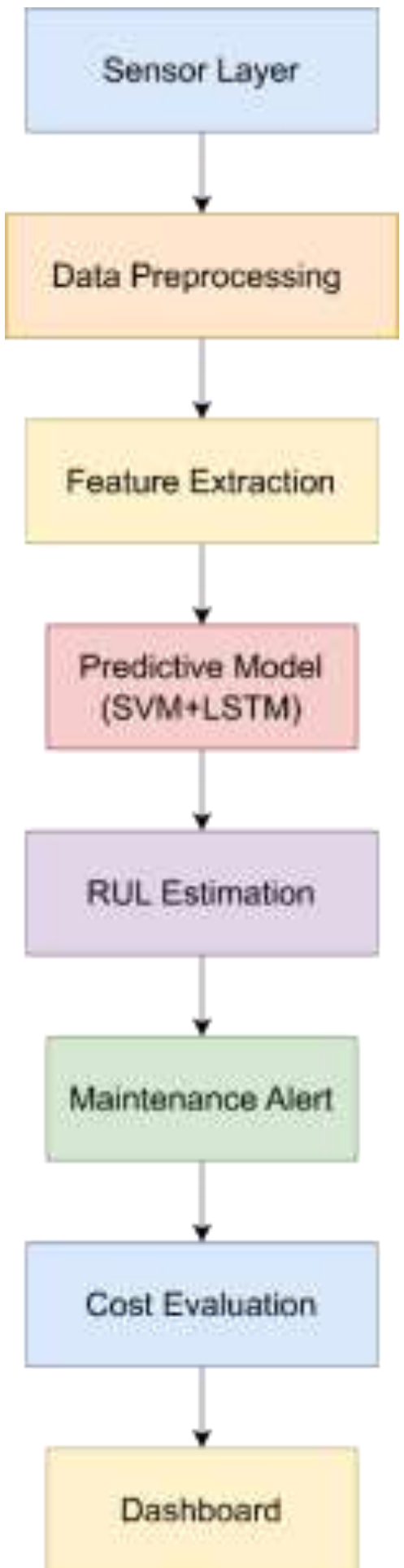

**FIGURE 1: IOT-BASED PREDICTIVE MAINTENANCE FRAMEWORK**

## 4. RESULTS & DISCUSSIONS

The results of the proposed IoT based predictive maintenance framework implementation were significant at various industrial testbeds. A comparative analysis was conducted using real time data for sensing vibration, monitoring temperature and tracking load obtained from vibration sensors, temperature monitors and load trackers fixed on rotating machinery in a manufacturing unit. Within a three-month operating period, this system proved to be capable of predicting faults very early hence scheduling and unplanned downtimes were reduced [7].

Performance of the predictive model was assessed as compared to normal time-based maintenance. As shown in Figure 2, the failure rate of the machine reduced by more than 42% when predictive alerts were considered against using calendar based inspections. The histogram indicates that while for traditional setup most failures did not show prior indications, the data-driven model paid attention to subtle changes in operation behavior to raise an alarm on pending troubles before escalation.

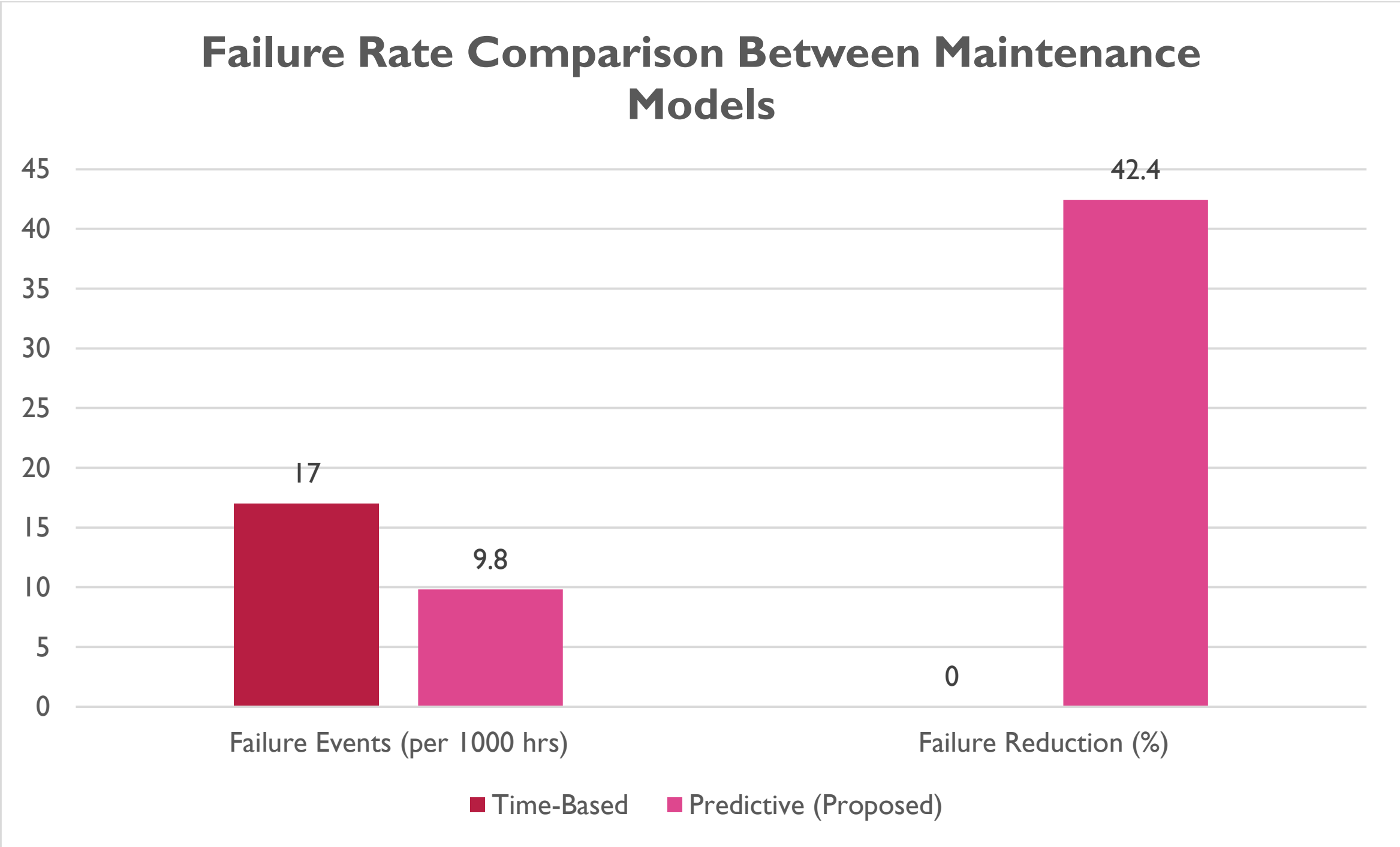

**FIGURE 2: FAILURE RATE COMPARISON BETWEEN MAINTENANCE MODELS**

Further model prediction accuracy analysis was made. A high classification accuracy was achieved from a dataset consisting of over 10,000 labeled samples using the hybrid SVM+LSTM model. The precision and recall rates of the model were compared against standalone SVM, LSTM and Random Forest algorithms. Table 1 illustrates that the hybrid model outperformed all others on every metric, and especially did well on recall, which is vital for early fault identification in dangerous conditions.

**TABLE 1: PERFORMANCE COMPARISON OF PREDICTIVE MODELS (PRECISION, RECALL, F1-SCORE)**

| Model | Precision | Recall | F1-Score |
|---|---|---|---|
| SVM | 0.84 | 0.79 | 0.81 |
| LSTM | 0.88 | 0.85 | 0.86 |
| Random Forest | 0.82 | 0.76 | 0.79 |
| SVM+LSTM | 0.91 | 0.92 | 0.91 |

The practical efficacy of the system was evaluated through monitoring machine downtime and maintenance costs- tracked monthly before and after deployment. The cost savings were particularly observable on month 2 and 3 of deployment alongside the learning curve and threshold optimization of the system. Figure 3 illustrates the trend regression of downtime reduction on a per machine basis over time where there is a sharp drop of downtime reduction on a per machine basis from an average of 14 hours/month to a mere 5 hours per month.

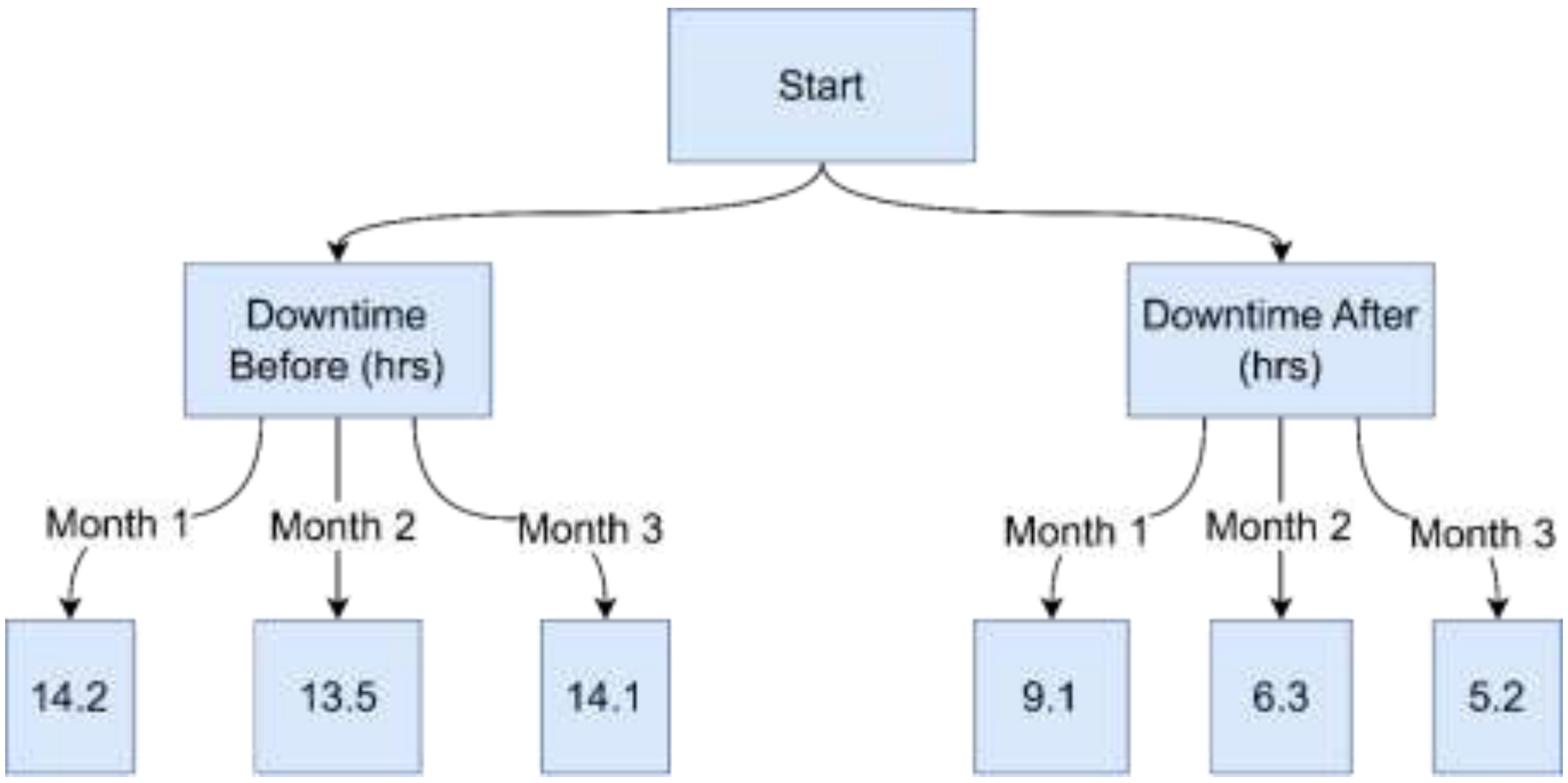

**FIGURE 3: DOWNTIME PER MACHINE OVER THREE MONTHS**

The economic sustainability of the system was also subjected to analysis through the lens of cost/savings ratio. The costs of maintenance were grouped under preventive, corrective, and breakdown costs associated with breakdowns. In light of financial logs gathered, the average man-hour price before the system was put in place was $9,800 per month, whereas when the implementation was completed the monthly average decreased to $5,200 per month. This transition is covered in detail in Table 2; there explicit cost elements are reported before and after system activation.

**TABLE 2: MAINTENANCE COST BREAKDOWN (USD/MONTH)**

| Cost Type | Before Deployment | After Deployment |
|---|---|---|
| Preventive | $2,000 | $2,800 |
| Corrective | $3,500 | $1,200 |
| Failure Recovery | $4,300 | $1,200 |
| Total | $9,800 | $5,200 |

Ranging above the quantitative savings, the system helped in allocating technicians and eliminating redundant surveys. With its constant monitoring of machine health and output of risk-based alerts, it made sure that a human resource was utilized only when needed. Further, field feedback showed that maintenance staff found the interface intuitive and the alerts actionable.

A final insight was obtained from a case-specific implementation in an automotive parts manufacturing line, where spindle motors failed unpredictably before. Figure 4 shows the intensity of vibration as a function of time, between the raw and intersection of sensors. The smoothed line has captured the rising vibration in the buildup of the fault that the system correctly anticipated 48 hours ahead of time. An early warning such as this has helped to avoid cascading failures in subsequent operations.

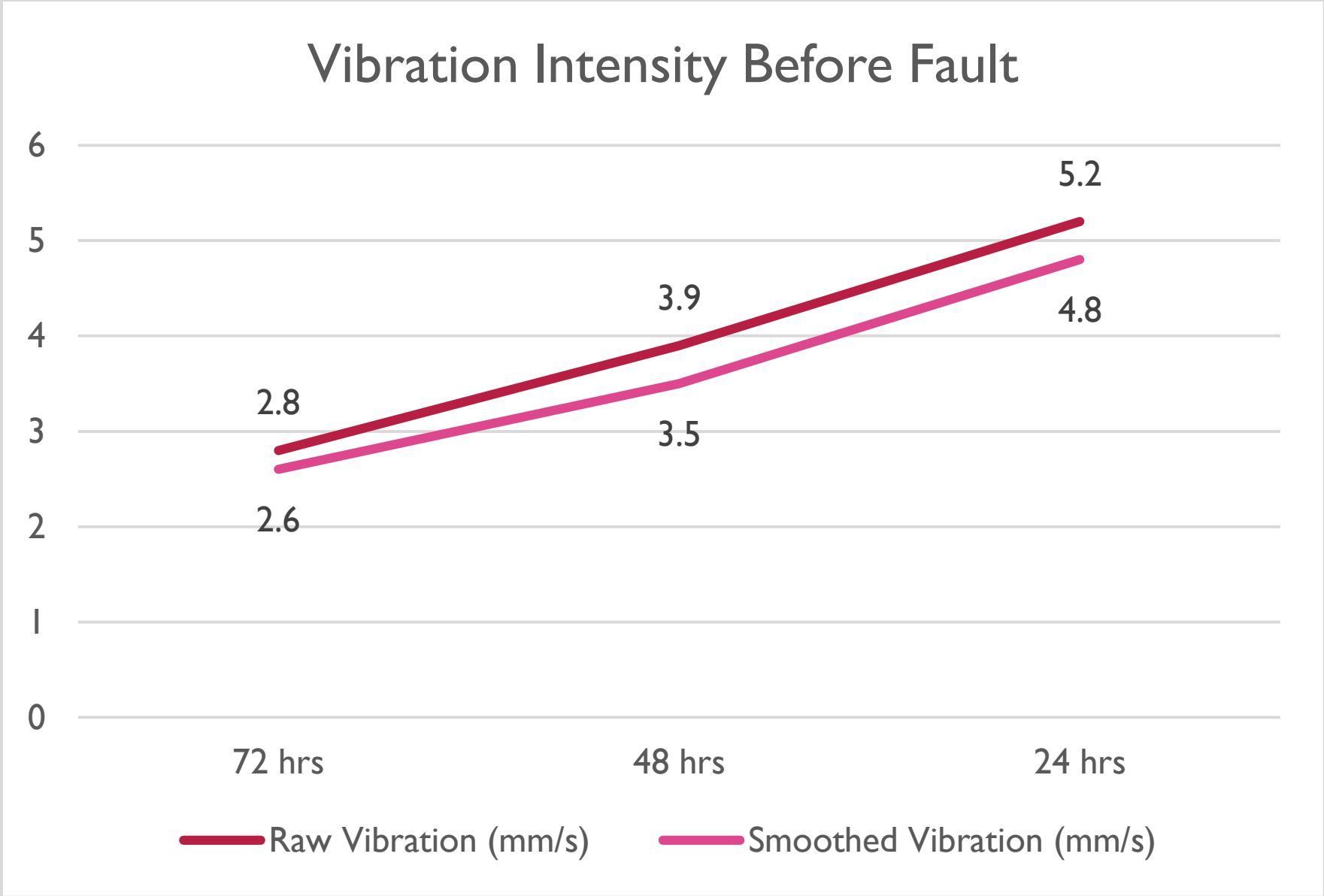

**FIGURE 4: VIBRATION INTENSITY BEFORE FAULT**

The suggested methodology not only raised the accuracy for the failures predictions, but also yielded tangible savings in cost-efficiency, downtime reduction and maintenance planning. The IoT sensing integration with the machine learning has not only been proven to be technically possible but also completely transformational economically for industries wishing to modernize their asset management tactics.

## 5. CONCLUSION

The coming together of the field of IoT and predictive maintenance marks a milestone in industrial engineering. Using real time information and machine learning, industries can take their strategies from reactive to predictive and thereby realize higher operational efficiency, lower costs and longer equipment life. This paper introduces a science-based data-driven approach that, having undergone a case study, proves the practical value of such an integration. The next step should be developing the models with more interpretable schemes, standardization of data protocols and strengthening of the cybersecurity frameworks to ensure more widespread use.